\documentclass[A4paper, 12pt, twoside]{article}
\usepackage{fancyhdr}
\usepackage{anysize}
\usepackage{graphicx}
\usepackage{amsmath}
\usepackage{amsfonts}
\usepackage[utf8]{inputenc}
\usepackage[colorlinks]{hyperref}
\usepackage{tensor}
\usepackage{epsfig}

\marginsize{1.5cm}{1.5cm}{0cm}{1.5cm}
\newcommand{\calP}{\mathcal{P}} 
\newcommand{\varZ}{\zeta}

\date{}
\title{One-Loop Partition Function,\\Gauge Accessibility and Spectra in AdS$_3$ Gravity}
\author{Joel Acosta$^{\dag}$, Alan Garbarz$^{\ddag}$, Andr\'es Goya$^{\mathsection}$, Mauricio Leston$^{\mathsection}$}

\begin{document}
\maketitle
\vspace{.25cm}
\begin{minipage}{.9\textwidth}
    \small \it 
	\begin{center}
    $^\dag$ Departamento de  Matem\'atica-FCEN-UBA \& IMAS-CONICET,
	Ciudad Universitaria, Pabell\'on 1, 1428, Buenos Aires, Argentina.
     \end{center}
\end{minipage}

\vspace{.25cm}
\begin{minipage}{.9\textwidth}
    \small \it 
	\begin{center}
    $^\ddag$ Departamento de F\'isica-FCEN-UBA \& IFIBA-CONICET,
	Ciudad Universitaria, Pabell\'on 1, 1428, Buenos Aires, Argentina.
     \end{center}
\end{minipage}

\vspace{.25cm}
\begin{minipage}{.9\textwidth}
    \small \it 
    \begin{center}
    $^\mathsection$ Instituto de Astronom\'ia y F\'isica del Espacio (IAFE),
	Pabell\'on IAFE-CONICET, Ciudad Universitaria, C.C. 67 Suc. 28, Buenos Aires, Argentina.
     \end{center}
\end{minipage}
\vspace{.5cm}

\begin{abstract}
\small
We continue the study of the one-loop partition function of AdS$_3$ gravity with focus on the square-integrability condition on the fluctuating fields. In a previous work we found that the Brown-Henneaux boundary conditions follow directly from the $L^2$ condition. Here we rederive the partition function as a ratio of Laplacian determinants by performing a suitable decomposition of the metric fluctuations. We pay special attention to the asymptotics of the fields appearing in the partition function. We also show that in the usual computation using ghost fields for the de Donder gauge, such gauge condition is accessible precisely for square-integrable ghost fields. Finally, we compute the spectrum of the relevant Laplacians in thermal AdS$_3$, in particular noticing that there are no isolated eigenvalues, only essential spectrum. This last result supports the analytic continuation approach of David, Gaberdiel and Gopakumar. The purely essential spectra found are consistent with the independent results of Lee and Delay of the essential spectrum of the TT rank-2 tensor Lichnerowickz Laplacian on asymptotically hyperbolic spaces.
\vspace{.5cm}

\begin{flushleft}
\hrulefill\\
\footnotesize
{E-mails: jacosta@dm.uba.ar, alan@df.uba.ar, agoya@iafe.uba.ar, mauricio@iafe.uba.ar}
\end{flushleft}

\end{abstract}

\newpage
\section{Introduction}

It is a major goal in high energy theoretical physics to fully understand the partition function of AdS$_3$ quantum gravity. It encodes the spectrum of the putative quantum theory and can be used to gain insight about a possible dual CFT. In \cite{Giombi:One.Loop}, Giombi et al. took an important step by computing from the gravitational theory the one-loop partition function $Z^{\text{(1-loop)}}(\tau,\bar{\tau})$ on a solid torus. More precisely,  they departed from the quadratic action around thermal AdS$_3$, chose a gauge, introduced ghosts fields,  and by the heat-kernel method computed the relevant functional determinants in the solid torus $\mathbb{H}^3/\Gamma$. Here $\Gamma$ implements the appropriate quotient which we will explain in more detail later. They landed at the following result,
\begin{equation}\label{Z1loop}
    Z^{\text{(1-loop)}}(\tau,\bar{\tau})=\prod_{m=2}^\infty \frac{1}{|1-q^m|^2},\qquad q:=e^{2\pi i \tau} \,,
\end{equation}
where $\tau$ is the modular parameter of the (conformal) boundary torus. Such a partition function is the product of two  characters over an irreducible representation of the Virasoro algebra. This was a  robust indicative of the existence of an underlying two-dimensional CFT. This fact has a classical precursor: Brown and Henneaux in \cite{Brown.Henneaux:AdS3} showed that, with sensible asymptotic  boundary conditions on  \textit{Lorentzian} metrics in General Relativity with negative cosmological constant, the algebra of conserved charges is nothing but two copies of the Virasoro algebra with central charge $c=\frac{3\ell}{2 G}$. It is worth noticing that in \cite{Giombi:One.Loop} the expectation and confirmation of the appearance of the Virasoro algebra in the partition function did not came from the imposition of Brown-Henneaux boundary conditions, but from an argument borrowed by from \cite{Maloney.Witten}. 

In 2007 Maloney and Witten had already argued that if Brown-Henneaux boundary conditions are imposed, the computation of the gravity partition function as  a trace of $e^{-\beta H-i \theta J}$ should give a Virasoro character. In short, the reason is that with Brown-Henneaux asymptotic boundary conditions the physically different gravitational configurations are organized in Virasoro coadjoint representations \cite{Garbarz.Leston:1403, Barnich:2014zoa}. Coming back to the work \cite{Giombi:One.Loop}, we see that although it was reasonable to expect that a Virasoro character would show up in the Euclidean computation of the partition function, the Brown-Henneaux boundary conditions were not really imposed at any moment. And moreover, we can wonder how is it that a Virasoro character makes its appearance by a purely Euclidean path-integral computation.

Motivated by the question of where are implicitly assumed/imposed the asymptotic boundary conditions of Brown-Henneaux in the path-integral of \cite{Giombi:One.Loop}, we found in \cite{Acosta:2020} that the square integrability of the metric perturbations directly imply the (Euclidean version of) Brown-Henneaux boundary conditions. Even more, it was possible to make a distinction, within the space of asymptotic symmetry generators, between proper asymptotic vectors (those square integrable, which are trivial under the coadjoint representation) and improper asymptotic vectors (those which are not square integrable, implying that are not trivial under the coadjoint representation, modulo proper ones). This distinction is essentially the same as that of \cite{Brown.Henneaux:AdS3}.

However, a few but important points were left to be clarified in our previous work. First of all, the ghost fields in \cite{Giombi:One.Loop} appear from the start in the path integral but is not a priori clear if they should belong to the space of $L^2$ vectors or to the space of vectors which generate $L^2$ metric perturbations. This issue is relevant because of the following: if the space of vector fields is too small, it might not include the proper asymptotic vectors. On the contrary, if it is too big, it might exceed the domain of the operator whose determinant appears in the partition function (see \cite{Castro:2017mfj} for a discussion on these possibilities and the implications on partition functions).  

Second, and relevant for any gauge theory, is the matter of the accessibility of the gauge condition: can the de Donder gauge be reached from any $L^2$ perturbation by an infinitesimal diffeomorphism? This is crucial since ensures that the Faddeev-Popov method actually works and it is also needed in order to implement a ``constraint first'' quantization procedure. As we will see, this is closely related to the question about the space to which the ghost fields belong.

In third place, we have the issue of describing in detail the spectrum of the relevant operators entering in the 1-loop partition function. This is crucial for understanding the computation of the functional determinant of such operators. In order to be as clear as possible, let us first introduce these operators. The result \eqref{Z1loop} of \cite{Giombi:One.Loop} comes from the expression of the 1-loop partition function as a quotient of functional determinants,
\begin{equation}
    Z^{\text{(1-loop)}}(\tau,\bar{\tau})=\frac{\text{det} \left(-{\Delta}^{(1)}\right)}{\sqrt{\text{det} \left(-{\Delta}^{(2)}\right)\text{det} \left(-{\Delta}^{(0)}\right)}} \,,
\end{equation}
where $\Delta^{(i)}$ are Laplacian-like operators with different masses acting on tensor fields of rank $i$. The operator $\Delta^{(2)}$ acts on traceless-symmetric tensors. In \cite{Giombi:One.Loop} the determinants were computed using a combination of the heat-kernel method for $\mathbb{H}^3$ and the method of images to find the heat kernel in $\mathbb{H}^3/\Gamma$. It is hard to say if this procedure manages to capture the full spectrum of the operators in $\mathbb{H}^3/\Gamma$ when it starts from the essential spectrum in $\mathbb{H}^3$. It is known in the mathematical literature that the spectrum of Lichnerowicz Laplacians on $\mathbb{H}^3$ is a  purely  essential spectrum \cite{Lee2001math5046L, DELAY200233}. In order to simplify the previous expression, it is possible to work with one less operator, as showed for example in \cite{David:2009xg},
\begin{equation}\label{Z1loopdeterminats}
    Z^{\text{(1-loop)}}(\tau,\bar{\tau})=\frac{\text{det}^{1/2}_T\left( -\nabla^{2}+2\right)}{\text{det}^{1/2}_{TT} \left(-\nabla^{2}-2\right)} \,.
\end{equation}
Here the determinant in the numerator is computed over transverse vector fields, while the one in the denominator is computed over transverse-traceless symmetric rank-2 tensors. In \cite{David:2009xg} the heat kernels in $\mathbb{H}^3/\Gamma$ are obtained from a clever analytic continuation from the heat kernel on a quotient of $S^3$. Again, it is hard to say whether this procedure captures the full spectrum of the operators, or may be it misses some possible isolated points in the spectrum.  

Although there are available computations related to the spectrum of the operators in \eqref{Z1loopdeterminats} for the BTZ  \cite{Datta:2011za}, and as we mentioned the heat kernel of thermal AdS3 was computed in \cite{David:2009xg} and \cite{Giombi:One.Loop}, to the best of our knowledge there is no known result of the \textit{full} spectrum of the relevant operators in Thermal AdS$_3$. In the mathematics community it has been studied in great depth the essential spectrum of Laplacian-like operators in asymptotically hyperbolic manifolds in arbitrary dimensions. In particular, in \cite{Lee2001math5046L} and \cite{DELAY200233} it was shown that the  essential spectrum of the Lichnerowicz Laplacian of a symmetric-traceless rank-2 tensor is the ray $[(n-1)(n-9)/4,\infty)$. Also they showed that for $\mathbb{H}^n$ this is the whole spectrum. For $n=3$ this dimensions implies that $-\nabla^{2}-2$ has essential spectrum $[1,\infty)$, for any asymptotically hyperbolic 3-manifold.  What is not clear if there is a non-empty discrete spectrum, in other words isolated eigenvalues. A similar  situation holds for vector fields \cite{Donnelly81,Donnelly84}.
 
Having this in mind, in this paper we address the following questions:
\begin{enumerate}
    \item What are the sensible boundary conditions on the ghost fields?
    \item Is de Donder gauge accessible with proper asymptotic diffs?
    \item What is the complete spectrum of the relevant operators in thermal AdS$_3$? 
\end{enumerate}

These questions guide the organization of this work. In Section 2 we show that the ghost fields must belong to $L^2$ by scrutinizing the differential operators acting on a convenient decomposition of the metric fluctuations inside the path integral. In addition, we show that indeed de Donder gauge is accessible by gauge transformations generated by $L^2$ vector fields. This is another way of seeing that ghost fields must belong to $L^2$. Finally, in Section 3 we compute the full spectrum of the relevant operator in the path integral in thermal AdS$_3$, following and improving \cite{Datta:2011za}.
\section{The Partition Function}

\subsection{Identifications and Coordinates of Hyperbolic Spaces}

\begin{figure}[ht]
    \centering
    \includegraphics[width=0.5\textwidth]{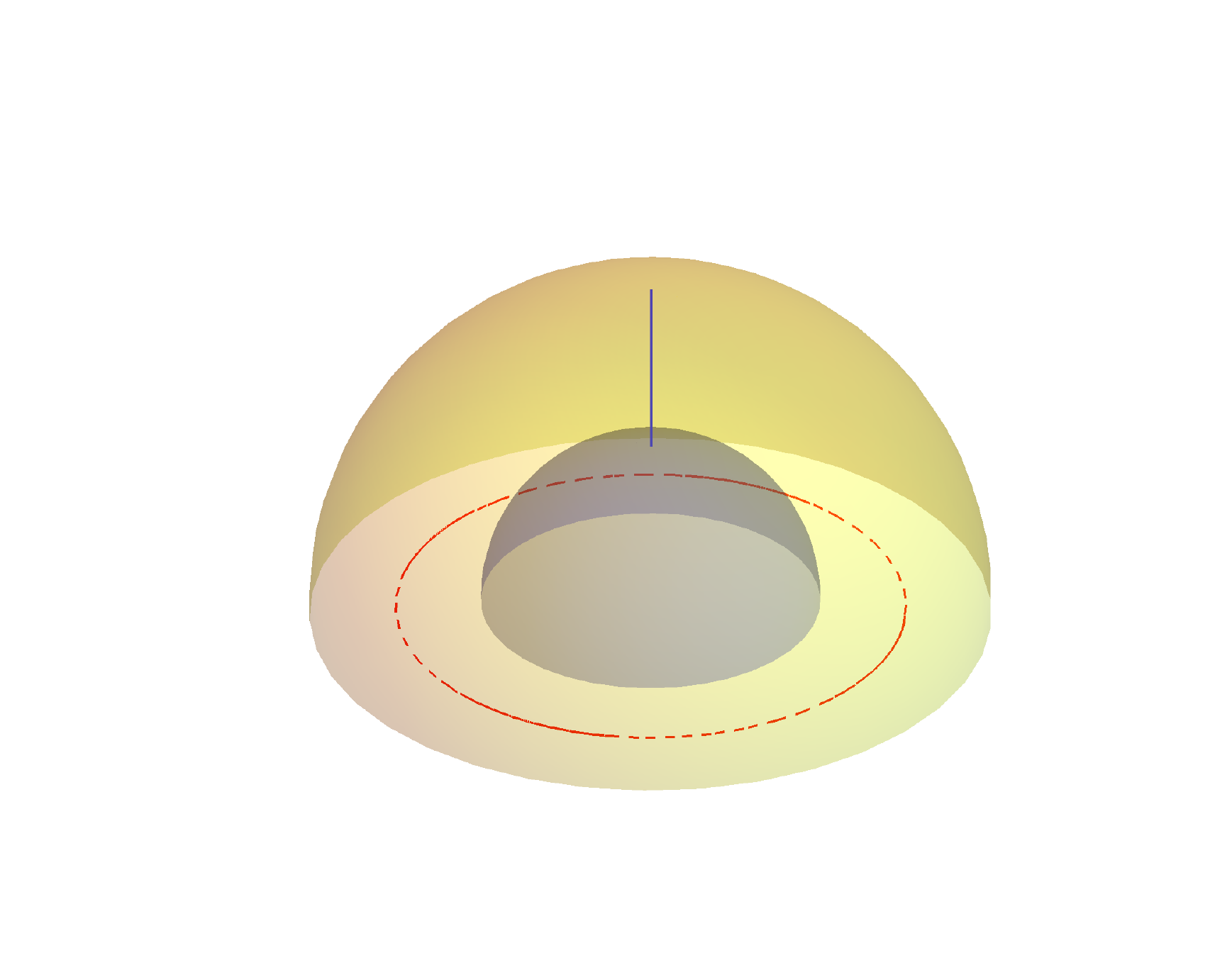}
    \caption{The fundamental region of $\mathbb{H}^3$/$\Gamma$ lies between the two domes. The surface of the inner dome is identified with the surface of the exterior one. The red line indicates the contractible cycle and the blue line indicates the non-contractible one.}
    \label{fig:cupula}
\end{figure}

The hyperbolic space $\mathbb{H}^3$ can be described by the half-space model as $(X,Y,Z)\in \mathbb{R}^3$ with $Z>0$ and with metric 
\begin{equation}
\label{eq:metrica_hiperbolica}
ds^2=\frac{dX^2+dY^2+dZ^2}{Z^2} \,.
\end{equation}
We are going to consider the quotient $\mathbb{H}^3/\Gamma$ where $\Gamma$ is the group generated by an isometry $\gamma$ which combines a rotation in the $X,Y$ plane by a fixed amount $\theta$ and a global dilation by a factor $e^{\beta}$

\begin{equation}
\label{eq:identificacion}
    \begin{pmatrix}
    X \\ Y \\ Z
    \end{pmatrix} \sim \gamma \cdot \begin{pmatrix}
    X \\ Y \\ Z
    \end{pmatrix} = e^{\beta}  \begin{pmatrix}
    \cos(\theta)X - \sin(\theta)Y \\ \sin(\theta)X + \cos(\theta)Y \\ Z
    \end{pmatrix} \,.
\end{equation}
This identifies the semispheres of radius $1$ with the semisphere of radius $e^{\beta}$, $e^{2\beta}$ and so on. We are going to consider the fundamental region inside the semispheres of radius $1$ and $e^{\beta}$ (see Figure \eqref{fig:cupula}). Usually,  the parameters $\theta$ and $\beta$ are combined into a new one $\tau=\frac{1}{2\pi}(\theta+i\beta$). The quotient $\mathbb{H}^3/\Gamma$ can be thought of as a solid torus where $\tau$ is the modular parameter of its conformal boundary.

Performing the following change in the coordinates of $\mathbb{H}^3$ 
\begin{equation}
    \begin{aligned}
        \lambda&=\sqrt{\dfrac{X^2+Y^2}{Z^2}} \,,\\
        \psi&=\frac{1}{2}\log\left(X^2+Y^2+Z^2\right) \,,\\
        \varphi&=\arctan\left(Y/X\right) \,,
    \end{aligned}
\end{equation}
we arrive to the familar (Euclidean) AdS$_3$  metric

\begin{equation}
    ds^2 = \dfrac{1}{\left(\rho^2+1\right)}d\rho^2 + \left(\rho^2+1\right)d\psi^2 + \rho^2d\varphi^2 \,,
\end{equation}
here $\varphi$ is the $2\pi$-periodic angular coordinate. The identification (\ref{eq:identificacion}) acts on the coordinates as 

\begin{equation}
    (\psi\,,\,\varphi)\sim (\psi\,,\,\varphi +2\pi)\sim (\psi + \beta\,,\, \varphi + \theta) \,,
\end{equation}
On the other hand, performing the following change in the coordinates 

\begin{equation}
\label{eq:Transformations}
    \begin{array}{rcccl}
     \tanh^2(\varZ)&=&\dfrac{r^2-r_+^2}{r^2+r_-^2}&=&\dfrac{X^2+Y^2}{X^2+Y^2+Z^2} \,, \\
     x_{-}&=&r_+\tau+r_{-}\phi&=&\arctan\left(Y/X\right) \,, \\
     x_{+}&=&r_{+}\phi-r_{-}\tau &=&\dfrac{1}{2}\log\left(X^2+Y^2+Z^2\right) \,,
    \end{array}
\end{equation}
we obtain the metric
\begin{equation}
\label{eq:BTZ_metrica}
\begin{aligned}
    ds^2&=d\varZ^2 +\sinh^2(\varZ)dx_-^2 + \cosh^2(\varZ)dx_{+}^{2}\\
    &=\dfrac{r^2}{\left(r^2-r_+^2\right)\left(r^2+r^2_-\right)}dr^2 + \dfrac{\left(r^2-r_+^2\right)\left(r^2+r_-^2\right)}{r^2} d\tau^2 + r^2\left(d\phi-\dfrac{r_+r_-}{r^2}d\tau\right)^2 \,, \\
    \end{aligned}
\end{equation}
which corresponds to the Euclidean version of the so-called BTZ black hole \cite{BTZ, BTZ:Geometry}. The $(x_+,x_-)$ coordinates are now identified as
\begin{equation}
    (x_+\,,\,x_-)\sim(x_+ \,,\, x_- +2\pi) \sim (x_+ +\beta \,,\, x_- +\theta) \,.
\end{equation}
 The parameters $\theta$ and $\beta$ can be expressed in terms of $r_+$ and $r_-$ by writing $(\tau,\phi)$ in terms of $x_{\pm}$ and imposing the periodicity condition $\phi\sim \phi + 2\pi$. We obtain  $\beta=2\pi r_+$, $\theta=2\pi r_-$ and the periodicity of the coordinates become

\begin{equation}
\begin{aligned}
    (x_+\, ,\, x_-)&\sim(x_+ \,,\, x_- +2\pi) \sim (x_+ +2\pi r_+ \,,\, x_- +  2\pi r_-) \,, \\
    (\tau\,,\,\phi)&\sim (\tau\,,\,\phi+2\pi) \sim \left(\tau+\frac{2\pi r_+}{r_+^2+r_-^2}\,,\,\phi+\frac{2\pi r_-}{r_+^2+r_-^2}\right) \,.
\end{aligned}
\label{eq:periodicidades}
\end{equation}
To visualize the coordinates $(r,x_-,x_+)$ in the half-space model it is useful to invert the transformations (\ref{eq:Transformations})

\begin{equation}
    \begin{aligned}
        X&=\left(\dfrac{r^2-r_+^2}{r^2+r_-^2}\right)^{1/2}\cos(x_-)e^{x_+} \,, \\
        Y&=\left(\dfrac{r^2-r_+^2}{r^2+r_-^2}\right)^{1/2}\sin(x_-)e^{x_+} \,, \\
        Z&=\left(\dfrac{r_+^2+r_-^2}{r^2+r_-^2}\right)^{1/2}e^{x_+} \,.
    \end{aligned}
\end{equation}
For a fixed $r$ these are the equations of cones through the origin, which means $r$ can be thought of as a zenithal angle. The coordinates $x_-$ and $x_+$ are coordinates on those cones, acting as angular and radial coordinates respectively. 

After the identification, the intersections between the $r$-constant cones and the fundamental region are tori. Any such torus serves as a boundary dividing two regions inside the fundamental region, one inner and one reaching the conformal boundary. Because of this, an $r-$constant torus can be used to evaluate boundary terms by computing them at a fixed value of $r$ and then taking limit $r\to \infty$ (see \cite{Takhtajan:2002cc} for a broader discussion). One might be tempted do this procedure by taking horizontal planes at a fixed value of $Z$ and then take the limit $Z\to 0$. However, due to the identification, the surfaces of constant $Z$ are not the boundaries of any region inside the fundamental region. Put differently, constant $Z$ surfaces are not invariant under the quotient.

\subsection{Decomposition of Metric Perturbations}
Here we describe the decomposition that we use for the rank-2 tensors in order to evaluate the action. It is customary to split the perturbations in a transverse-traceless component, a longitudinal component and a tracefull component. Following \cite{Deser} we further impose the trace component to be transverse, this cannot be done in general but it is possible when the background spacetime is maximally symmetric. 
This extra condition guarantees the orthogonality between the three components and allows us to rewrite the action in a very convenient form.

We are going to consider tensors that are square integrable with respect to the  inner product
\begin{equation}
     \langle  T \mid T' \rangle = \int d^3x \sqrt{g}\, \bar{T}\cdot T' \,,
     \label{eq:Producto interno}
\end{equation}
where $\bar{T}\cdot T'$ refers to the full contraction with respect of the metric $g$. Thus, for example, for rank-2 tensors the inner product is
\begin{equation}
     \langle  T \mid T' \rangle = \int d^3x \sqrt{g}\, \bar{T}^{\mu \nu} T'_{\mu \nu} = \int d^3x \sqrt{g}\,  g^{\mu \alpha} g^{\nu \beta}\bar{T}_{\mu \nu} T'_{\alpha \beta} \,.
\end{equation}

A square-integrable symmetric tensor $T_{\mu \nu}$ can be decomposed into a transverse and a longitudinal part as follows
\begin{equation}
T_{\mu \nu} = T^t_{\mu \nu} + \left(\nabla_{\mu} V_{\nu} +\nabla_{\nu}V_{\mu}\right) =T^t_{\mu\nu} + \mathcal{L}_{\mu\nu}V\,,  \quad \text{with}  \quad \nabla^{\mu} T^{t}_{\mu \nu}=0 \,,
\label{eq:decomp_T}
\end{equation}
where the vector $V$ is also square integrable with respect to (\ref{eq:Producto interno}). The divergence of the original tensor is a square-integrable vector and it is carried by the longitudinal part
\begin{equation}
    \nabla^{\nu} T_{\mu\nu} = B_{\mu} =\nabla^{\nu} \left(\nabla_{\mu} V_{\nu} +\nabla_{\nu}V_{\mu}\right) \,.
    \label{eq:invert_V}
\end{equation}
By integrating by parts it can be shown that the operator that takes $V_{\mu}$ into $B_{\mu}$ is negative-semi-definite

\begin{equation}
    \int dx^3\sqrt{g} V^{\nu}\nabla^{\mu}(\nabla_{\nu}V_{\mu}+\nabla_{\mu}V_{\nu})= - \dfrac{1}{2}\int dx^3 \sqrt{g}(\nabla_{\nu}V_{\mu}+\nabla_{\mu}V_{\nu})^2 \leq 0 \,,
\end{equation}
where the boundary term vanishes if the vectors are square integrable. Thus, it can be inverted and we can always find the vector $V_{\mu}$ of the decomposition (\ref{eq:decomp_T}).

Now, we want to decompose the transverse part $T^{t}_{\mu\nu}$ into a traceless and a tracefull part while keeping the transversality property. This can be done by defining 
\begin{equation}
\label{eq:descomp_chi}
    T^{t}_{\mu \nu}= T^{TT}_{\mu \nu} + \dfrac{1}{2}\left(g_{\mu\nu}\nabla^2-\nabla_{\mu}\nabla_{\nu}\right)\chi -g_{\mu\nu}\chi = T^{TT}_{\mu\nu} + \Theta_{\mu\nu}\chi \,,
\end{equation}
where $\Theta_{\mu\nu}\chi$ is given by
\begin{equation}
\label{eq:Ttracefull}
    \Theta_{\mu\nu}\chi = \dfrac{1}{2}\left(g_{\mu\nu}\nabla^2-\nabla_{\mu}\nabla_{\nu}\right)\chi \,,
\end{equation}
and $\chi$ is a square-integrable function. Taking trace in the equation (\ref{eq:descomp_chi}) we find that the relevant operator is
\begin{equation}
   g^{\mu\nu}T^{t}_{\mu\nu}=T^t= (\nabla^{2}-3)\chi \,,
   \label{eq:invert_chi}
\end{equation}
where the trace $T^t$ is a square-integrable function. By integrating by parts again, we can see that this operator is also negative semi-definite. Therefore, any square-integrable symmetric tensor can be decomposed as

\begin{equation}
    T_{\mu \nu}= T^{TT}_{\mu \nu} + \Theta_{\mu\nu}\chi + \mathcal{L}_{\mu\nu} V = T_{TT} + T_{\Theta} + T_{L} \,,
    \label{eq:descomp_h}
\end{equation}
where we denote the transverse-traceless part as $T_{TT}$, the transverse-tracefull part by $T_{\Theta}$, and the longitudinal part by $T_{L}$. To find the vector field $V$ and the scalar field $\chi$ one has to invert the operators appearing in (\ref{eq:invert_V}) and (\ref{eq:invert_chi}) respectively. The orthogonality between the components in (\ref{eq:descomp_h}) follows from the respective transverse and traceless properties when implemented in the inner product (\ref{eq:Producto interno}).

\subsection{Computation of the Partition Function}
In this section we proceed to compute the one-loop partition function  around AdS$_3$
\begin{equation}
    Z = \int Dh\, e^{-S^{(2)}\left[h\right]} \,,
\end{equation}
where $S^{(2)}$ is the Einstein-Hilbert action expanded to quadratic order in the perturbation $h$ around the Euclidean AdS$_3$ background, i.e., $\bar{g} = g + h$ with $g$ the background metric. From now on, all contractions, raising and lowering of indices, and inner products such as \eqref{eq:Producto interno} are taken with respect to the background metric $g$. The quadratic action, fixing $\Lambda=-1$ and taking into account the Gibbons-Hawking term, is \cite{Witten:2018lgb}
\begin{equation}
\begin{aligned}
\label{eq:action}
    S^{(2)}=S&=-\int d^3x\sqrt{g}\left(\dfrac{1}{4}h_{\mu\nu}(\nabla^2+2)h^{\mu\nu} - \dfrac{1}{8}h\nabla^2 h +\dfrac{1}{2}G(h)^2\right) \,. \\
\end{aligned}
\end{equation}
The last term in the above expression invites to define the de Donder gauge functional
\begin{equation}
    G_{\nu}(h) = \nabla^{\mu}h_{\mu\nu} - \dfrac{1}{2}\nabla_{\nu}h \,.
    \label{eq:Gauge}
\end{equation}
As expected, the quadratic action (\ref{eq:action}) is invariant under linearized diffeomorphisms, i.e., $h_{\mu\nu}\to h_{\mu\nu}+\mathcal{L}_{\xi}g_{\mu\nu}$, for $h$ and $\xi$ in $L^2$. As a result of this and the orthogonality properties of the decomposition (\ref{eq:descomp_h}), (\ref{eq:action}) factorizes into two independent pieces and no longer  depends on the longitudinal component, that is
\begin{equation}
    S[h_{\mu \nu}]=S[ h_{TT}+ h_{\Theta} + h_{L}] = S[ h_{TT}]+ S[h_{\Theta}] \,.
        \label{eq:S_split}
\end{equation}

The explicit form of the actions can be easily written as quadratic operators acting of each component as follows

\begin{alignat}{2}
\label{eq:S_htt}
    S[h_{TT}]&= \int d^3x \sqrt{g} \left( \dfrac{1}{4}(h_{TT})_{\mu\nu}\left(-\nabla^2-2\right)(h_{TT})^{\mu\nu}\right) \,, \\
    S[h_{\Theta}]&= -\int d^3x\sqrt{g} \left(\dfrac{1}{8}\chi\left(-\nabla^2+3\right) \left(-\nabla^2+2\right)^2\chi\right) \,.
\end{alignat}

We see that the action for the trace mode is negative definite. This is a well-known problem for the Euclidean action \cite{Gibbons.Hawking.Indefiniteness} and the standard procedure to deal with it is to rotate the contour of integration to be parallel to the imaginary axis. 
The partition function can be easily computed using the decomposition (\ref{eq:descomp_h}) and the property (\ref{eq:S_split})

\begin{equation}
\begin{aligned}
    Z&=\int Dh\, e^{-S\left[h\right]}  =\int Dh_{TT}Dh_{V}Dh_{\Theta}\, e^{-S\left[h_{TT}+h_{V}+h_{\Theta}\right]} \\ 
        &= \int Dh_{TT}DVD\chi\,J_V J_{\chi}e^{-S\left[h_{TT}+h_{\Theta}(\chi) + \mathcal{L}_Vg \right ]} \\
        &=J_V J_{\chi} \left(\int DV\right)\left(\int Dh_{TT}\, e^{-S[h_{TT}]}\right) \left(\int D\chi\, e^{-S[h_{\Theta}(\chi))]}\right) \,,
\end{aligned}
\end{equation}
where  $J_V$ and $J_{\chi}$ are the Jacobians associated to the change in the integration variables $h_{V}\to V$ and $h_{\Theta}\to \chi$ respectively. . The integral over the vectors diverges and it is nothing but the volume of the space of square-integrable vectors which can be identified with proper diffeomorphims \cite{Acosta:2020}. The last two integrals are Gaussian and they can be written in terms of the determinants of the operators appearing in (\ref{eq:S_htt}). The explicit computation of the Jacobians  $J_V$ and $J_{\chi}$ is carried out in the  appendix (\ref{apendice_jacobianos}). Collecting everything together we obtain 

\begin{equation}
    Z=\dfrac{\text{det}^{1/2}_{V_T}\left(-\nabla^2+2\right)}{\text{det}^{1/2}_{h_{TT}}\left(-\nabla^2-2\right)} \,,
    \label{eq:Z}
\end{equation}
where the operator in the numerator acts on square-integrable transverse vectors and the operator in the denominator on square-integrable transverse-traceless rank-2 tensors. This is precisely the result of \cite{Yasuda:1983hk} using the de Donder gauge-fixing condition.

\subsubsection{Accessibility of the Gauge Choice}
Even though we have not explicitly chosen a gauge in the previous computation, let us discuss briefly what would have happened if we had chosen one.   The usual gauge-fixing procedure consists in picking a gauge condition $G(h)=0$ and, using the Faddeev-Popov procedure, write an identity operator as follows 

\begin{equation}
\label{eq:Fadeev_popov}
    1 = J_{\eta}\int D\eta \, \delta(G(h+\mathcal{L}_\eta g)) \,,
\end{equation}
were $J_{\eta}$ is the Jacobian associated to the change of variables from the metric $h$ to the ghost vector field $\eta$. The next step is to insert (\ref{eq:Fadeev_popov}) in the path integral with the objective of factorize the gauge redundancy. In order to do this, the previous expression has to be valid for all square-integrable perturbations $h$. That is, the gauge has to be accessible, which means that for every square-integrable $h$ there must exist a perturbation $h_0$ which fulfills the gauge condition and a vector $\eta$ such that
\begin{equation}
    h_{\mu\nu}=h_0 + \mathcal{L}_{\eta}g \,.
\end{equation}
Using the decomposition (\ref{eq:descomp_h}) for $h$, the de Donder gauge condition (\ref{eq:Gauge}) turns out to be
\begin{equation}
    G(h^{TT}_{\mu\nu}) + G(\Theta_{\mu\nu}\chi) + G(\mathcal{L}_{\mu\nu}V) = G(h_0)+ G(\mathcal{L}_{\mu\nu}\eta) \,.
\end{equation}
Given that $h^{TT}$ and $h_0$ fulfill the gauge condition, the above expression is simply given by
\begin{equation}
    -\dfrac{1}{2}\nabla^{\nu}\left(\nabla^2-3\right)\chi - P^{\nu}(V) = - P^{\nu}(\eta) \,,
\end{equation}
where $P^{\nu}$ is defined by
\begin{equation}
    P^{\nu}(\eta)= \left(-\nabla^2+2\right)\eta^{\nu}\,.
    \label{eq:accesibilidad}
\end{equation}

In order to guarantee the accessibility of the gauge choice we need to be able to solve \eqref{eq:accesibilidad} for the auxiliary ghost vector field $\eta$. Notice that $P$  is definite positive when acting on square-integrable vectors and therefore it can, at least formally, be inverted\footnote{In Section \ref{sec:ProofSpectra} we prove that $0$ does not belong to the spectrum of $P$ and therefore $P$ is invertible.}. Thus, because the two terms in the LHS of (\ref{eq:accesibilidad}) are square integrable, it suffices to demand the ghost fields to belong to the space of $L^2$ vector fields and not to the space of vectors which generate $L^2$ metric perturbations.
\section{Spectrum of Laplacian Operators}

In this section we are going to  analyze the $\mathbb{H}^3/\Gamma$ spectrum of the operators appearing in ($\ref{eq:Z}$) acting on rank-2 symmetrical transverse-traceless tensors and transverse vector fields respectively. The spectra of these operators are fully understood in $\mathbb{H}^3$, for example, it is a well-known fact that the full spectrum of the Lichnerowicz Laplacian $\Delta_{LL}$ acting on rank-2 symmetrical traceless tensors in $\mathbb{H}^3$ is $[-3,\infty)$ with no eigenvalues below it \cite{DELAY200233,delayTT}. The Lichnerowicz Laplacian in this case is related with the ordinary Laplacian by \cite{Gibbons.Perry}
\begin{equation}
    \left(\Delta_{LL}+4\right)T_{\mu\nu}=\left(-\nabla^2 - 2\right)T_{\mu\nu} \,,
\end{equation}
which means the spectrum of the operator $(-\nabla^2-2)$ is $[1,+\infty)$. If the space is asymptotically hyperbolic (as it is $\mathbb{H}^3/\Gamma$) with dimension $n=3$, then $[-3,\infty)$ is the essential spectrum of the Lichnerowicz Laplacian and there might exist some eigenvalues below $-3$. A similar scenario holds for the case of the vector fields \cite{Donnelly81,Donnelly84}.

In order to study the spectra of the operators we are going to use an auxiliary first-order differential equation and prove that their spectra are closely related. We use coordinates $(z,x_+,x_-)$ which are related to the coordinates described in (\ref{eq:Transformations}) by 
\begin{equation}
    z=\tanh^2(\varZ) = \dfrac{r^2-r_+^2}{r^2+r_-^2} \,.
\end{equation}
In these coordinates the metric is
\begin{equation}
\begin{aligned}
    ds^2&= d\varZ^2 + \sinh^2(\varZ)dx^2_{-} + \cosh^2(\varZ)dx^2_{+} \\
         &=\dfrac{1}{4z(1-z)^2} dz^2 + \dfrac{z}{1-z}dx^2_{-} + \dfrac{1}{1-z}dx^2_{+} \,,
\end{aligned}
\label{eq:coordenadas}
\end{equation}
where the $x_+$ and $x_-$ have the periodicity \eqref{eq:periodicidades}. In these coordinates it turns out that the components of the solutions of the differential equations can be written in terms of the boundary components $\{ \xi_{+}\,, \xi_{-} \}$ for vectors, and $\{ h_{++} \,, h_{--} \,, h_{+-} \}$ for the rank-2 tensors. The resulting differential equations can be solved with hypergeometric functions. As usual, we are going to impose a delta-like normalization condition upon the solutions to analyze the spectra of the operators\footnote{Notice that it is possible to obtain $L^2$ functions by multiplying by a bump-like function with an appropriate localized support in the interior of the space and as close to  the asymptotic region as one wants \cite{HislopSigal}}.

\subsection{Rank-1 Tensor Field}

\subsubsection{Vector Laplacian and its Equivalent First-Order Operator}
Our goal is to determine the spectra of the operators appearing in (\ref{eq:Z}) subject to tranversality conditions. We will start with  the rank-1 case in which we have
\begin{equation}
\label{eq:laplaciano_vectores}
    \left(-\nabla^2 +2\right) \xi_{\mu} = \alpha^2 \xi_{\mu} \,,
\end{equation}
with the constraint
\begin{equation}
\label{eq:transverse_vectores}
	\nabla^{\mu}\xi_{\mu} = 0\,.
\end{equation}
Let us notice that the above system of differential equations can be formally factorized in terms of first-order differential operators as follows
\begin{equation}\label{PP}
	m^2\calP_{\mu}^{\nu}(-m)\calP_{\nu}^{\rho}(m)\xi_{\rho}=(\nabla^2+2+m^2)\xi_{\mu} \,,
\end{equation}
where $\calP(\pm m)$ is defined by\footnote{Since we will find later on that $m$ belongs to the spectrum of  $\epsilon\indices{_\mu^\nu^\rho} \nabla_{\nu}$ if $m \in \mathbb{R}-\left\{0\right\}$, and the spectrum is a closed set of the complex plane, $m=0$ necessarily belongs to the spectrum of $\epsilon\indices{_\mu^\nu^\rho} \nabla_{\nu}$. So we do not need to consider the case $m=0$ independently.}\textsuperscript{,}\footnote{We denote the Levi-Civita symbol by $\varepsilon_{\mu\nu\rho}$ and the Levi-Civita pseudo tensor by $\epsilon_{\mu\nu\rho}=\sqrt{g}\varepsilon_{\mu\nu\rho}$.}
\begin{equation}
	\calP_{\mu}^{\nu}(\pm m)\xi_{\nu}= \left(\delta_{\mu}^{\nu} \pm \dfrac{1}{m}\epsilon\indices{_\mu^\rho^\nu}\nabla_{\rho}\right)\xi_{\nu} \,,
    \label{Poperator}
\end{equation}
and $m$ is given by
\begin{equation}
	m^2= \alpha^2 - 4 \,.
\end{equation}

The factorization from above suggests that, instead of solving the second-order problem of \eqref{eq:laplaciano_vectores}, we might solve the first-order one given by\footnote{It is common to find in the literature that this is a shortcut first shown in  \cite{Tyutin:1997yn}. However, as there are unbounded operators involved, it seems to be more complicated than just finding the eigenvalues of $\calP$. For example, it is obvious that an eigenfunction of $\calP(m)$ with zero eigenvalue must be an eigenfunction of \eqref{eq:laplaciano_vectores} by the identity \eqref{Poperator}. However, how do we know that the spectrum of the second order differential operator comes completely from that of the first order differential operator? It will turn out to be the case, but as far as we know, it has not been analyzed at all in the present context.}
\begin{equation}
\label{eq:primer_orden_vectores}
	\epsilon\indices{_\mu^\nu^\rho} \nabla_{\nu}\xi_{\rho} = - m \xi_{\mu} \,,
\end{equation}
or, in terms of the operator $\calP(\pm m)$,
\begin{equation}
\label{eq:primer_orden_vectoresP}
	\calP_{\mu}^{\nu}(\pm m)\xi_{\nu} = 0 \,.
\end{equation}
As we said before, we will prove that if one obtains the spectra of \eqref{eq:primer_orden_vectoresP} then one automatically obtains the spectra of \eqref{eq:laplaciano_vectores}. 

As a first step, we will show that the eigenfunctions of \eqref{eq:primer_orden_vectores} are eigenfunctions of \eqref{eq:laplaciano_vectores} too. Notice that taking derivative in both sides of (\ref{eq:primer_orden_vectores}) we obtain
\begin{equation}
\begin{aligned}
    -m\nabla_{\mu}\xi^{\mu}= \epsilon\indices{^\mu^\nu^\rho}\nabla_{\mu}\nabla_{\nu} \xi_{\rho}&= \dfrac{1}{2}\epsilon\indices{^\mu^\nu^\rho}\left[\nabla_{\mu},\nabla_{\nu}\right]\xi_{\rho} = -\dfrac{1}{2}\epsilon\indices{^\mu^\nu^\rho} R\indices{^\lambda_\rho_\mu_\nu}\xi_{\lambda} \\
    & = -\dfrac{1}{2}\epsilon\indices{^\mu^\nu^\rho} \left(g_{\lambda\nu}g_{\rho\mu}- g_{\lambda\mu}g_{\rho\nu}\right)\xi^{\lambda}=0\,,
\end{aligned}
\end{equation}
therefore, solutions of \eqref{eq:primer_orden_vectores} are automatically transverse.

To recover the second-order differential equation \eqref{eq:laplaciano_vectores} we just need to replace in (\ref{eq:primer_orden_vectores}) a solution $\xi$ of the equation itself. What we obtain is
\begin{equation}
\label{eq:eps_eps_xi}
\begin{aligned}
    m^2\xi_{\mu}&= \epsilon\indices{_\mu^\nu^\rho}\epsilon\indices{_\rho^\alpha^\beta}\nabla_{\nu}\nabla_{\alpha}\xi_{\beta} = \left(g_{\mu\alpha}g_{\nu\beta} - g_{\mu\beta}g_{\nu\alpha}\right) \nabla^{\nu}\nabla^{\alpha}\xi^{\beta} = \nabla_{\beta}\nabla_{\mu}\xi^{\beta} - \nabla^2\xi_{\mu} \\
    &=\nabla_{\mu}\nabla_{\beta}\xi^{\beta} + \left[\nabla_{\beta},\nabla_{\mu}\right]\xi^{\beta} - \nabla^2\xi_{\mu} =\left(- \nabla^2- 2\right)\xi_{\mu} \,,
\end{aligned}
\end{equation}
where we have used the transverse condition \eqref{eq:transverse_vectores} in the last step. Therefore, the eigenfunctions of \eqref{eq:primer_orden_vectores} also satisfy
\begin{equation}
    \left(-\nabla^2+2\right)\xi_{\mu}= \left(4+m^2\right)\xi_{\mu}\,,
    \label{eq:P_funcion_m}
\end{equation}
which is the second-order equation (\ref{eq:laplaciano_vectores}) with $\alpha^2=4+m^2$. 

The above computations give us a hint that we might find the spectra of \eqref{eq:laplaciano_vectores} by finding the spectra of \eqref{eq:primer_orden_vectores} instead. However, knowing the eigenfunctions of $\calP$ are eigenfunctions of the second-order operator in \eqref{eq:laplaciano_vectores} is not enough. We can ask ourselves if the converse holds (and it does by a simple argument). But the difficult question is whether the essential spectrum (the part of the spectrum that are not isolated eigenvalues) of $\calP$ is enough to reconstruct the essential spectrum of the operator in \eqref{eq:laplaciano_vectores}. In the following subsection we will show that this is indeed the case by relating both spectra.

\subsubsection{Proof of the Relation between the Spectra of \texorpdfstring{$\calP(\pm m)$}{P(+/-m)} and \texorpdfstring{$-\nabla^2-2$}{-Delta-2}}
\label{sec:ProofSpectra}
Some clarifications related to the computations performed in the last section are in place. First, we had always implicitly consider the self-adjoint extensions of the differential operators. Second, we are going to use repeatedly that $\calP(m)\calP(-m) = \calP(-m)\calP(m)$, which holds since we can think that the dense domain where the LHS and RHS are originally defined is that of  compactly supported vector fields,
which is a core for the Laplacian \cite{Lee.Fredholm}. Therefore, taking closure on both sides gives self-adjoint operators, which must coincide since self-adjoint extensions are unique. 

Starting with the isolated eigenvalues, we can say that if $m^2$ is an isolated eigenvalue of $-\nabla^2-2$, then both square roots $\pm |m| $ are eigenvalues of $\epsilon\indices{_\mu^\nu^\rho} \nabla_{\nu}$. This is because if  $\calP(m)\calP(-m)\phi_n=0$ holds  with finite multiplicity for some vectors $\phi_n$, it follows that if $\phi$ is one of these eigenvectors associated to $m^2$, then $\calP(\pm m)\phi$ is an eigenvector of $\calP(\mp m)$, implying that both $\pm m$ are eigenvalues of $\epsilon\indices{_\mu^\nu^\rho} \nabla_{\nu}$. Conversely, if $m$ is an eigenvalue of $\epsilon\indices{_\mu^\nu^\rho} \nabla_{\nu}$, by (\ref{PP}) $m^2$  is an eigenvalue of $-\nabla^2-2$. This means that if we characterize in detail the spectrum of the first order differential operator $\epsilon\indices{_\mu^\nu^\rho} \nabla_{\nu}$, we have a full characterization of the spectrum of the original operator $-\nabla^2-2$. However, so far we have analyzed the discrete spectrum, it remains to do the same for the essential spectrum to complete the reasoning. 

Let us now show that if $m^2 \in \mathbb{R}$ is in the essential spectrum of $-\nabla^2-2$, then one of its square roots $\pm |m|$ is in the essential spectrum of $\epsilon\indices{_\mu^\nu^\rho} \nabla_{\nu}$. Conversely, for any $m$ in the essential spectrum of $\epsilon\indices{_\mu^\nu^\rho} \nabla_{\nu}$, $m^2$ belongs to the essential spectrum of  $-\nabla^2-2$. 

First we need to recall a few facts about spectral theory of self-adjoint operators, which we denote as $A$ \cite{HislopSigal}, and Dom$(A)$ refers to its domain:
\begin{enumerate}
    \item $\lambda \in $ Spec$(A) $ if and only if  $\exists \left\{\phi_n\right\} \subset $ Dom$(A), ||\phi_n||=1, $ such that $\lim_{n\rightarrow \infty}||(A-\lambda)\phi_n||=0$
    \item If the previous sequence $ \left\{\phi_n\right\} \subset $ Dom$(A)$ satisfies that $\lim_{n\rightarrow \infty} \langle \psi,\phi_n \rangle=0$ for any $\psi \in L^2$, it is called a Weyl sequence and $\lambda \in \text{Spec}_{\text{ess.}}(A)$, the essential spectrum of $A$. Conversely, if $\lambda \in \text{Spec}_{\text{ess.}}(A)$ there exists a corresponding Weyl sequence.  This is the Weyl criterion.
    \item The remaining sequences of item 1., namely those which are not Weyl sequences (do not have a zero weak limit) correspond to a $\lambda$ in the discrete spectrum and  have finite degeneracy.  
\end{enumerate}

What the second statement is telling us is that $\lambda$ is in the essential spectrum if and only if there is a Weyl sequence associated to it. We then have to show that if we have a Weyl sequence associated to $m^2$ for the operator $-\nabla^2-2$, then we have a Weyl sequence for $\calP(m)$ or $\calP(-m)$ and $\lambda=0$ and vice versa.  

Let us start and assume that there is such Weyl sequence of vector fields $\left\{\phi_n\right\}$, so we have 
\[||( -\nabla^2-2 - m^2)\phi_n|| \rightarrow 0 \,, \] 
which by (\ref{PP}) is equivalent to 
\[||\calP(m)\calP(-m)\phi_n|| \rightarrow 0 \,. \] 
We are going to show next that one can choose either $\left\{\phi_n\right\}$ or  $\left\{\frac{\calP(-m)\phi_n}{||\calP(-m)\phi_{n} ||}\right\}$ as Weyl sequence for  $\lambda=0$ and $\calP(-m)$ or $\calP(m)$ respectively, depending on whether the limit $\lim_{n \rightarrow \infty}||\calP(-m)\phi_n||$ is zero or not. Let us first assume that $\lim_{n \rightarrow \infty}||\calP(-m)\phi_n|| = 0$. Then the original Weyl sequence $\left\{\phi_n\right\}$ is by hypothesis a Weyl sequence for $\calP(-m)$ and $\lambda=0$. Namely, $m$ belongs to the spectrum of $\epsilon\indices{_\mu^\nu^\rho} \nabla_{\nu}$.

Now, let us consider  the case in which $\lim_{n \rightarrow \infty}||\calP(-m)\phi_n||=k\neq{0}$. The vectors  $\psi_n:=\frac{\calP(-m)\phi_n}{||\calP(-m)\phi_{n} ||}$ are normalized vectors and have a weak zero limit since\footnote{If $\lim_{n \rightarrow \infty}||\calP(-m)\phi_n||$ does not exist, then there is a subsequence $\psi_{n_k}:=\frac{\calP(-m)\phi_{n_k}}{||\calP(-m)\phi_{n_k} ||}$ which satisfies $\epsilon< ||\psi_n||$  for some $\epsilon > 0$ and by the same arguments it is a Weyl sequence for $\calP(m)$ and $\lambda=0$. }:

$$ \lim_{n\rightarrow{\infty}}\langle\psi,\psi_n\rangle=
\lim_{n\rightarrow{\infty}}\langle\calP(-m)\psi, \frac{1}{||\calP(-m)\phi_{n} ||}\phi_n\rangle=1/k \lim_{n\rightarrow{\infty}}\langle\calP(-m)\psi, \phi_n\rangle=0 \,. $$
It remains to show that $\lim_{n\rightarrow{\infty}}||\calP(m)\psi_n||=0$ but that follows from the assumption that $\left\{\phi_n\right\}$ is a Weyl sequence for $\calP(m)\calP(-m)$. 

The converse also holds . This comes from another way of characterizing/defining the essential spectrum of a self-adjoint operator: $\lambda$ is in the essential spectrum of $A$ if and only if $A-\lambda \mathbb{I}$ has either infinite dimensional kernel or cokernel \cite{EdmundsEvans}.

Using this characterization of the essential spectrum, if we assume that 0 in the essential spectrum of $\calP(-m)$, then $\calP(-m)$ has either infinite dimensional kernel or cokernel. We have to show that the same holds for $\calP(m)\calP(-m)$. Let us suppose that dim Ker$\calP(-m)=\infty$, then $\calP(m)\calP(-m)$ has infinite dimensional kernel as well. On the other hand, if dim Coker$\calP(-m)=\infty$, since Ran$(\calP(-m)\calP(m))\subseteq$ Ran$(\calP(-m))$, we have
\[\text{dim Coker}(\calP(m)\calP(-m))=\text{dim Coker}(\calP(-m)\calP(m))\ge \text{dim Coker}(\calP(-m))=\infty \,, \] 
which implies that dim Coker$(\calP(m)\calP(-m))=\infty$. We have called Ran$(\calP(-m))$ to the range of $\calP(-m)$. 

With the discussion above,  we are now certain that by studying the spectrum of the operators $\calP(m)$ for any $m$, we have complete control over the spectrum of $-\nabla^2-2$. It is also important to note that it did not matter the tensorial structure of the operators, so we can apply the same logic later to analyze the case of the rank-2 tensor.

\subsubsection{Solutions to the First-Order Differential Operator and its Essential Spectrum}
In the previous section we showed that the spectrum of $-\nabla^2-2$ can be found by solving the spectrum of a simpler  first-order  differential operator. Now, we will proceed to solve $\calP(m)\xi = 0$ and find its eigenfunctions for different values of  $m$.

It will be useful to rewrite the equation \eqref{eq:primer_orden_vectores} for 1-forms using coordinates (\ref{eq:coordenadas})
\begin{equation}
    \begin{aligned}
    \epsilon[\xi]_{\mu} &= \epsilon\indices{_\mu^\nu^\rho} \nabla_{\nu} \xi_{\rho} = -m \xi_{\mu} \\
    &=-\frac{1}{2z}(\partial_+ \xi_+ - \partial_- \xi_-) dz - 2z(z-1)(\partial_+ \xi_z - \partial_z \xi_+) dx_- -2(z-1)(\partial_z\xi_- - \partial_-\xi_z)dx_+ \,.
    \end{aligned}
    \label{eq:eps_vector}
\end{equation}
The solutions of this equation also have to satisfy the equation (\ref{eq:eps_eps_xi}).  Writing down the components of the Laplacian acting on 1-forms, it can be shown that
\begin{equation}
    \begin{aligned}
    (\nabla^2\xi)_{-} -\Delta \xi_-  +2 \epsilon[\xi]_{+} = -2 \xi_-\,,\\
    (\nabla^2\xi)_{+} - \Delta \xi_{+} - 2 \epsilon[\xi]_{-} = -2 \xi_+\,,\\
    \end{aligned}
\end{equation}
where $\Delta$ is the scalar Laplacian acting on the components of the 1-form. Using the fact that $\xi$ is a  solution of (\ref{eq:eps_vector}) the equation (\ref{eq:eps_eps_xi}) can be written in terms of the following matrix equation 
\begin{equation}
    \left(\nabla^2+2\right)\begin{pmatrix}\xi_- \\ \xi_+\end{pmatrix} = \Delta\begin{pmatrix}\xi_- \\ \xi_+\end{pmatrix} +
    \begin{pmatrix}
     0 & 2m \\-2m & 0 
    \end{pmatrix}\begin{pmatrix}\xi_- \\ \xi_+\end{pmatrix} = -m^2 \begin{pmatrix}\xi_- \\ \xi_+\end{pmatrix} \,.
\end{equation}
This can be diagonalized using  $\xi_{1,2}=1/2(\xi_+\mp i\xi_- )$ obtaining
\begin{equation}
\begin{pmatrix}
\Delta +(-2im +m^2) & 0 \\
        0 & \Delta +(2im +m^2)\\
\end{pmatrix}
\begin{pmatrix}
\xi_1 \\ \xi_2 \\
\end{pmatrix} = 0 \,.
\label{eq:lap_diag_vec}
\end{equation}
To solve this for $\xi_1$ and $\xi_2$ we use the ansatz
\begin{equation}
    \begin{aligned}
    \xi_1= e^{i(k_+ x^+ + k_- x^-)}z^{\alpha_1}(1-z)^{\beta_1}F_1(z)\,,\\
    \xi_2= e^{i(k_+ x^+ + k_- x^-)}z^{\alpha_2}(1-z)^{\beta_2}F_2(z)\,,
    \end{aligned}
    \label{eq:descomp_F}
\end{equation}
where $k_-$ has to be an integer and $k_+=(n-k_-r_-)/r_+$ with $n$ an integer due to the periodicity conditions (\ref{eq:periodicidades}). Explicitly, the differential equation for the functions $F_{1,2}(z)$ is given by
\begin{equation}
\begin{aligned}
    \left[z(1-z)\dfrac{\partial^2 F_i}{\partial z^2} +( (2\alpha_i+1)-z(2\alpha_i + 2\beta_i +1))\dfrac{\partial F_i}{\partial z} - \left(\frac{k_+^2}{4}+(\alpha_i+\beta_i)^2\right)F_i\right]\\ -\dfrac{1}{z}\left(\dfrac{k_-^2}{4}-\alpha_i^2\right)F_i + \dfrac{4\beta_i(\beta_i-1)+m(m-2i\epsilon_i)}{4(1-z)}F_i=0\,,
    \end{aligned}
\end{equation}
where $i=1,2$ and $(\epsilon_{1},\epsilon_2)=(1,-1)$. We see that it is enough to study the case $k_+\geq 0$ and $k_- \geq 0$. If we choose the coefficients of the $1/z$ and the $1/(1-z)$ terms to be zero, i.e. if we choose $\alpha_i$ and $\beta_i$ to be
\begin{equation}
    \alpha_{1,2}=\alpha= \dfrac{k_-}{2} \,, \qquad \beta_1= \left\{ 1+i\frac{m}{2}\  ;\ -i\frac{m}{2} \right\} \,, \qquad \beta_2 = \left\{1-i\frac{m}{2}\ ; \ i\frac{m}{2}\right\} \,,
    \label{eq:alpha_beta}
\end{equation}
the above differential equation becomes the hypergeometric differential equation

\begin{equation}
    z(1-z)\dfrac{\partial^2 F_i}{\partial z} + (c-(a_i+b_i+1)z)\dfrac{\partial F_i}{\partial z} - a_ib_iF_i=0\,,
\end{equation}
with
\begin{equation}
	c=2\alpha+1 = k_-+1\,, \qquad a_i=\frac{k_-}{2}+i\frac{k_+}{2} + \beta_i\,, \qquad b_i=\frac{k_-}{2}-i\frac{k_+}{2}+\beta_i\,.
\label{eq:abc}
\end{equation}
The solutions are  hypergeometric functions $_2F_1(a_i,b_i;c_i;z)$ or $F(a_i,b_i,c_i,z)$ for simplicity. It is important to note that $k_-$ is an integer because of the periodicity of $x_-$. This is a pathological case for the differential equation where the two usual solutions are not independent and a second independent solution can be constructed by Frobenius' method and it looks like  (see for example \cite{Olver}) 
\begin{equation}
    F(a,b,k_-+1,z)\log(z) - \sum_{i=0}^{k_-}a_i(-z)^{-i} + \sum_{i=0}^{\infty}b_i z^{i}\,.
    \label{eq:solucion_patologica}
\end{equation}
We are going to neglect these solutions because they make the norm of the vectors diverge at $z\sim 0$. For the solutions of the form $F(a,b,c,z)$ because of the transformation rules of the hypergeometric function, the two possibilities of $\beta$ yield to the same result. We are going to choose 
\begin{equation}
    \beta_1= -i\dfrac{m}{2} =\beta\,, \qquad \beta_2=-i\dfrac{m}{2}+1 = \beta+1\,.
\end{equation}
Then the solutions to the equations (\ref{eq:lap_diag_vec}) are
\begin{equation}
\label{eq:sol_hyper_vec}
    \begin{aligned}
    \xi_1&= Ae^{i(k_+ x^+ + k_- x^-)}z^{\alpha}(1-z)^{\beta}F(a,b,c,z)\,,\\
    \xi_2&= Be^{i(k_+ x^+ + k_- x^-)}z^{\alpha}(1-z)^{\beta+1}F(a+1,b+1,c,z)\,,
    \end{aligned}
\end{equation}
with
\begin{equation}
	\alpha=\frac{k_-}{2}\,, \qquad c=k_- + 1\,, \qquad a=\frac{k_-}{2}+i\frac{k_+}{2} + \beta\,, \qquad b=\frac{k_-}{2}-i\frac{k_+}{2} + \beta\,.
\end{equation}
The constants $A$ and $B$ are not independent. This is because $\xi_+$ and $\xi_-$ still have to satisfy the first order equation (\ref{eq:eps_vector}). The deduction of the relation is in the appendix (\ref{apendice_polarizacion}) and the result is
\begin{equation}
    \dfrac{B}{A}  = \dfrac{2\beta - k}{2\beta + \bar{k}}\,,
    \label{eq:polarizacion_vector}
\end{equation}
with $k=k_-+ik_+$. To obtain the $\xi_+$ and $\xi_-$ components we have to invert the relation that we used to diagonalize the matrix (\ref{eq:lap_diag_vec}). Then the $z$ component of the equation (\ref{eq:eps_vector}) allows us to find the $\xi_z$ component in terms of the two others
\begin{equation}
    \begin{aligned}
    \xi_-&= iAz^{\alpha}(1-z)^{\beta}e^{i(k_+ x^+ + k_- x^-)}\left[F(a,b,c,z)  - (1-z)\left(\dfrac{2\beta - k}{2\beta + \bar{k}}\right) F(a+1,b+1,c,z) \right]\,,\\
    \xi_+&= Az^{\alpha}(1-z)^{\beta}e^{i(k_+ x^+ + k_- x^-)}\left[F(a,b,c,z)  + (1-z)\left(\dfrac{2\beta - k}{2\beta + \bar{k}}\right) F(a+1,b+1,c,z) \right]\,,\\
    \xi_z&=\dfrac{1}{4\beta z} \left(k_+\xi_+ -k_-\xi_-\right) = -\dfrac{1}{2\beta z} (k\xi_2+\bar{k}\xi_1)\,.
    \end{aligned}
\end{equation}
For the above solutions the norm does not diverge for $z\simeq0$ therefore we only\footnote{It might seem, by looking at (\ref{eq:PI_vectores}), that if $\alpha=0$ (i.e. $k_-=0$) the contribution of the $\xi_-$ component gives a problem near $z=0$  in the norm, but it is not the case. If $k_-=0$, then  the constant (\ref{eq:polarizacion_vector}) becomes $1$ and the terms $\mathcal{O}(0)$ of the two hypergeometrics in $\xi_-$ cancel each other.} need to check the behavior for $z\simeq1$. It is easier to analyze the asymptotic behavior of these solutions using the coordinate $z=\tanh^2(\varZ)$ where the norm of the vectors is 
\begin{equation}
\begin{aligned}
    \langle \xi,\xi\rangle &=\int dz dx_- dx_+\left(\left|\xi_{z}\right|^2 z +  \dfrac{\left|\xi_-\right|^2}{(1-z)z}  + \dfrac{\left|\xi_+\right|^2}{(1-z)} \right)\,, \\
    &=\int d\varZ dx_- dx_+\left(\left|\xi_{\varZ}\right|^2\sinh(\xi)\cosh(\xi) +  \left|\xi_-\right|^2 \dfrac{\cosh(\xi)}{\sinh(\xi)} + \left|\xi_+\right|^2 \dfrac{\sinh(\xi)}{\cosh(\xi)}\right)\,.
    \end{aligned}
    \label{eq:PI_vectores}
\end{equation}
Because of the decomposition (\ref{eq:descomp_F}) the integral in $x_+$ and $x_-$ are deltas. The radial part of the integral can be divergent depending on the asymptotic behavior of the functions $\xi_1$ and $\xi_2$. Using the asymptotic behavior computed in (\ref{apendice_asintotica}) we obtain 

\begin{equation}
\begin{aligned}
    &\xi_1\sim c_1 e^{(-2-im)\varZ} + c_2 e^{im\varZ}\,,\\
    &\xi_2\sim c_3 e^{-im\varZ} +c_4 e^{(-2+im)\varZ}\,,\\
    &\xi_{z}\sim e^{-2\varZ}\left(k\xi_1 + \bar{k}\xi_2\right)\,.
\end{aligned}
\end{equation}
First, it is important to note that both choices of $\beta$ in (\ref{eq:alpha_beta}) give the same asymptotic behavior.  It is easy to see that if $\text{Im}(m)\neq 0$ this norm is infinite. The only case where the form $\xi_{\mu}$ can be delta-like normalizable is when $m\in \mathbb{R}$. Therefore, the equation (\ref{eq:P_funcion_m}) turns out to be
\begin{equation}
    (-\nabla^2+2)\xi_{\mu}= \alpha^{2} \xi_{\mu}= (m^{2}+4)\xi_{\mu}\,.
\end{equation}
This means that the spectrum of the operator $-\nabla^2+2$ is $[4,+\infty)$ with no eigenvalues below. It is consistent with the spectral function computed it $\cite{Camporesi.Higuchi}$ and with the spectrum of the Hodge Laplacian acting on forms computed in \cite{Donnelly81,Donnelly84}.

\subsection{Rank-2 Tensor Field: First-Order Differential Operator, its Solutions and the Essential Spectrum}
For the rank-2 tensor case, the equation we want to solve is
\begin{equation}
\label{eq:laplaciano_tensores}
	\left(-\nabla^2 -2\right) h_{\mu \nu} = \,\alpha^2 h_{\mu\nu}\,,
\end{equation}
subject to the constraints
\begin{equation}
    \begin{aligned}
    \nabla^{\mu}h_{\mu\nu}=&\,0\,,\\
    g^{\mu\nu}h_{\mu\nu}=&\,0\,,
    \end{aligned}    
\end{equation}
where $h_{\mu\nu}$ is symmetric. Analogously as in the vector case, it can be shown that for symmetric tensors the eigenfunctions and the spectrum of these equations are related to the eigenfunctions and the spectrum of the following first order equation (see proof in section \eqref{sec:ProofSpectra})
\begin{equation}
    \epsilon[h]_{\mu\sigma}= \epsilon\indices{_\mu^\nu^\rho} \nabla_{\nu} h_{\rho\sigma} = -m h_{\mu\sigma} \,.
    \label{eq:epsilon_h}
\end{equation}
It can be shown that the solutions of (\ref{eq:epsilon_h}) are transverse and traceless. Using (\ref{eq:epsilon_h}) a second time we can show that its solutions satisfy the original differential equation with $\alpha=m^2+1$,
\begin{equation}
     (-\nabla^2-2)h_{\mu\nu}=(1+m^2)h_{\mu\nu}\,.
     \label{eq:L_funcion_m}
\end{equation}

Looking at the $(z-)$ and  $(z+)$ components of the equation (\ref{eq:epsilon_h}) we have 
\begin{equation}
\label{eq:z_components}
\begin{aligned}
    \epsilon[h]_{z-}= -h_{z+} + \dfrac{1}{2z}\left(\dfrac{\partial h_{--}}{\partial x^{+}} - \dfrac{\partial h_{+-}}{\partial x^{-}}\right) = - m h_{z-}\,,\\
    \epsilon[h]_{z+}= h_{z-} + \dfrac{1}{2z}\left(\dfrac{\partial h_{+-}}{\partial x^{+}} - \dfrac{\partial h_{++}}{\partial x^{-}}\right) = - m h_{z+}\,.\\
    \end{aligned}
\end{equation}
These equations together with the traceless condition allow us to write all the components in terms of the boundary components $h_{++},\,h_{+-}, h_{--}$. With help of the equation (\ref{eq:epsilon_h}) we can write
\begin{equation}
    \begin{aligned}
    &(\nabla^2h)_{--} - \Delta h_{--} - 4 \epsilon[h]_{+-} &=& -4 h_{--} + 2h_{++}\,,\\
    &(\nabla^2h)_{+-} - \Delta h_{+-} +2\epsilon[h]_{--} - 2 \epsilon[h]_{++} &=& -6 h_{+-}\,,\\
    &(\nabla^2h)_{++} - \Delta h_{++} + 4 \epsilon[h]_{+-} &=& +2h_{--} - 4h_{++}\,,\\
    \end{aligned}
\end{equation}
where $\Delta$ stands for the scalar Laplacian acting on the tensor components. Using that $h_{\mu\nu}$ is solution of the first order equation we can write the following matrix equation
\begin{equation}
    \left(\nabla^2+3\right)\begin{pmatrix}h_{--} \\ h_{+-} \\h_{++}\end{pmatrix} = \Delta\begin{pmatrix}h_{--} \\ h_{+-} \\h_{++}\end{pmatrix} +
    \begin{pmatrix}
     -1 & -4m & 2 \\2m & -3 & -2m \\ 2 & 4m & -1
    \end{pmatrix}\begin{pmatrix}h_{--} \\ h_{+-} \\h_{++}\end{pmatrix} = -m^2 \begin{pmatrix}h_{--} \\ h_{+-} \\h_{++}\end{pmatrix}\,,
\end{equation}
which can be diagonalized with
\begin{equation}
    \begin{pmatrix}
        h_{0} \\ h_{-} \\h_{+}
    \end{pmatrix}= 
    \begin{pmatrix}
        1/2 & 0 & 1/2 \\ -1/4& i/2& 1/4& \\ -1/4 & -i/2 &1/4\\
    \end{pmatrix} 
    \begin{pmatrix}
        h_{--} \\ h_{+-} \\h_{++}
    \end{pmatrix}\,,
    \label{eq:cambio_base_h}
\end{equation}
to obtain
\begin{equation}
\begin{pmatrix}
    \Delta + m^2 +1 & 0 & 0 \\
     0 & \Delta  + (m-2i)^2+1& 0 \\
     0& 0 &\Delta +(m+2i)^2+1\\
\end{pmatrix}
\begin{pmatrix}
h_0 \\ h_- \\ h_+ \\
\end{pmatrix} = 0\,.
\end{equation}
The solutions to these equations are obtained in the same way that in the vector case. Each component has two equivalent elections for its constant $\beta_i$, choosing them properly we can write the final result,
\begin{equation}
\begin{aligned}
    h_{+}&=A_{+}z^{\alpha}(1-z)^{\beta}F(a,b,c;z)e^{i(k_+ x^+ + k_- x^-)}\,,\\
    h_{0}&=A_{0}z^{\alpha}(1-z)^{\beta+1}F(a+1,b+1,c;z)e^{i(k_+ x^+ + k_- x^-)}\,,\\
    h_{-}&=A_{-}z^{\alpha}(1-z)^{\beta+2}F(a+2,b+2,c;z)e^{i(k_+ x^+ + k_- x^-)}\,,
\end{aligned}
\label{eq:sol_hyper_h}
\end{equation}
with
\begin{equation}
    \begin{array}{rcccl}
    \alpha&=&\dfrac{k_-}{2}\,, & &\\
    \beta&=&\frac{im}{2}-\frac{1}{2}\,, & &\\
    c&=&2\alpha+1 &=& k_- + 1\,, \\
    a&=& \alpha +\beta + i\frac{k_{+}}{2}&=& \frac{k_-}{2}  + i\frac{k_{+}}{2}+\beta\,,\\
    b&=& \alpha +\beta - i\frac{k_{+}}{2}&=& \frac{k_-}{2} - i\frac{k_{+}}{2} +\beta \,.\\
    \end{array}
\end{equation}
The constants $A_0,\, A_+,\, A_-$ are not free and they have to fulfill the first order equation (\ref{eq:epsilon_h}). The relation between them is computed in the appendix (\ref{apendice_polarizacion_2}) and the result is

\begin{equation}
\label{eq:relacion_polarizacion}
    \dfrac{A_0}{2A_+}= \dfrac{2\beta-k}{2\beta+\bar{k}} \,; \qquad \dfrac{2A_-}{A_0}= \dfrac{2(\beta+1)-k}{2(\beta+1)+\bar{k}}\,,
\end{equation}
with $k=k_-+ik_+$. To write down the components of the tensor we have to invert (\ref{eq:cambio_base_h}) and use (\ref{eq:sol_hyper_h}) together with (\ref{eq:relacion_polarizacion}) to obtain the boundary components $h_{++},h_{+-},h_{--}$. Then, inverting (\ref{eq:z_components}) we can obtain the components $h_{z+}$ and $h_{z-}$. Finally, the traceless conditions gives the $h_{zz}$ component.

Asymptotically the norm of the tensors is
\begin{equation}
    \langle h,h\rangle \sim \int d\varZ dx_+ dx_-\left(\left|h_{++}e^{-\xi}\right|^2 + \left|h_{--}e^{-\xi}\right|^2 + 2 \left|h_{+-}e^{-\xi}\right|^2 +2\left|h_{\xi+}\right|^2 + 2\left|h_{\xi-}\right|^2 +\left|h_{\varZ\varZ}\right|^2\right)\,,
\end{equation}
and the asymptotic behaviors of the solutions are
\begin{equation}
    \begin{aligned}
    h_{+}(\xi)&\sim c_1e^{im\varZ}e^{-3\varZ} &+& \quad c_2e^{-im\varZ} e^{\varZ}\,,\\
    h_{0}(\xi) &\sim c_3e^{im\varZ}e^{-\varZ} &+&\quad c_4e^{-im\varZ} e^{-\varZ}\,,\\
    h_{-}(\xi) &\sim c_5e^{im\varZ} e^{\varZ} \quad&+&\quad c_6e^{-im\varZ} e^{-3\varZ}\,.
    \end{aligned}
\end{equation}
It is easy to see that as well as in the vector case, this norm is badly divergent if $m$ has an imaginary part. The equation (\ref{eq:L_funcion_m}) says that the spectrum of the operator $-\nabla^2-2$ acting on symmetric transverse traceless rank-2 tensors is $[1,+\infty)$ with no eigenvalues below. This is consistent with the the well-known results for the essential spectrum of the Lichnerowicz Laplacian on $\mathbb{H}^3$ and $\mathbb{H}^3/\Gamma$ computed in \cite{DELAY200233,delayTT}. It is also consistent with the spectral functions computed in \cite{Camporesi.Higuchi}. 

\section{Summary}
We have shown that vector (ghost) fields must belong to the space of square-integrable vectors fields rather than to the space of vectors fields which generate square-integrable metric perturbations. This distinction is important because it is precisely this condition that splits the space of generators of asymptotic symmetries into proper and improper asymptotic vectors \cite{Acosta:2020}. We also have shown that this condition is enough to guarantee the accessibility of the de Donder gauge, which means that the gauge-fixing procedure in the computation of the partition function only removes the redundancies generated by proper diffeomorphisms, leaving intact the metric perturbations with a Brown-Henneaux fall-off behavior.

Then we switched to study the operators that appear in the final form of the one-loop partition function. We have proven that the spectrum these operators can be studied by looking at the spectrum of first-order operators. We concluded that there are no strictly square-integrable eigenfunctions for these operators in $\mathbb{H}^3/\Gamma$. By imposing the solutions to be delta-like normalizable, we have found the full essential spectra of the operators which are, in fact, the full spectra. Our results are consistent with several spectra computed on $\mathbb{H}^3$ and $\mathbb{H}^3/\Gamma$ \cite{DELAY200233,Donnelly81,Donnelly84,delayTT}.

\section*{Acknowledgments}
The authors thank Gaston Giribet and Mart\'in Mereb for enlightening discussions. AG thanks Michael Anderson, Erwann Delay and Rafe Mazzeo for correspondence. This work was partially supported by grants PIP and PICT from CONICET and ANPCyT.
\appendix
\newpage
\section{Jacobians}

\label{apendice_jacobianos}
\subsection{The Computation of \texorpdfstring{$J_{\chi}$}{Jx}}
The Jacobian $J_{\chi}$ can be obtained by imposing the following normalization for the integral

\begin{equation}
    1=\int Dh_{\Theta}\  e^{-\langle h_{\Theta}\mid h_{\Theta}\rangle} \,.
\end{equation}
Using the definition of the transverse-tracefull part $h_{\Theta}$ \eqref{eq:Ttracefull} and performing a change in the integration variable we obtain 
\begin{equation}
    1=\int Dh_{\Theta}\  e^{-\langle h_{\Theta}\mid h_{\Theta}\rangle}= \int D\chi J_{\chi}\, e^{-\langle \Theta_{\mu\nu}\chi\mid\Theta_{\mu\nu}\chi\rangle}= J_{\chi}\int D\chi\, e^{-\frac{1}{2}\langle \chi\mid(-\nabla^2+3)(-\nabla^2+2)\chi\rangle} \,.
\end{equation}
In the last equality we rewrite the inner product as a quadratic operator acting on the field $\chi$. The boundary behaviour of $\chi$ guarantees that the boundary term vanishes. The integral is Gaussian and we can write it as a determinant. The result is
\begin{equation}
    J_{\chi}= \text{det}_{\chi}^{1/2}(-\nabla^2+3)\times \text{det}_{\chi}^{1/2}(-\nabla^2+2) \,.
\end{equation}

\subsection{The Computation of \texorpdfstring{$J_{V}$}{JV}}

The factor $J_{V}$ is computed in a similar way to $J_{\chi}$. We impose the normalization
\begin{equation}
    1 = \int Dh_L \, e^{-\langle h_L \mid h_L \rangle} = \int Dv J_V e^{-\langle \mathcal{L}_V\mid \mathcal{L}_V\rangle} = J_V \int Dv \, e^{-2 \langle V \mid (-\nabla^\mu \mathcal{L}_{\mu\nu})[V] \rangle } \,.
    \label{eq:j_v1}
\end{equation}
Now, we can decompose $V=V_T + V_{\varphi}$ where the component $V_T$ is transverse and $V_{\varphi}= {\rm grad}(\varphi)$. These two components are orthogonal to each other. Up to a vanishing boundary term, we can rewrite the exponent in the last term of (\ref{eq:j_v1}) as the sum of two quadratic operators acting solely on $V_T$ and $V_{\varphi}$
\begin{equation}
    \langle V \mid -\nabla^{\mu}\mathcal{L}_{\mu\nu} [V]\rangle = \langle V_T \mid \left(-\nabla^2+2\right)V_T\rangle + 2 \langle \varphi \mid \nabla^2 (\nabla^2-2) \varphi\rangle \,.
\end{equation}
Performing a change of variables from $V_{\varphi}$ to $\varphi$ with  $J_{\varphi}$ the corresponding Jacobian, we obtain
\begin{equation}
    1 = J_V \left(\int D{V_T}e^{-2 \langle V_T\mid \left(-\nabla^2+2\right)V_T\rangle}\right) \left(D\varphi J_{\varphi} e^{-4 \langle \varphi \mid \nabla^2(\nabla^2-2) \varphi\rangle} \right) \,,
\end{equation}
where both integrals are Gaussian. The Jacobian $J_{\varphi}$  can be easily computed with the same procedure as before giving
\begin{equation}
 J_{\varphi} = \text{det}^{1/2}_{\varphi}(-\nabla^2) \,.
\end{equation}
Collecting everything together we finally obtain
\begin{equation}
    J_V= \dfrac{\text{det}^{1/2}_{V_T}(-\nabla^2+2)\times \text{det}^{1/2}_{\varphi}(-\nabla^2)\times \text{det}^{1/2}_{\varphi}(-\nabla^2+2)}{\text{det}^{1/2}_{\varphi}(-\nabla^2)} =  \text{det}^{1/2}_{V_T}(-\nabla^2+2)\times \text{det}^{1/2}_{\varphi}(-\nabla^2+2) \,.
\end{equation}

\section{The Polarization Constants}

\subsection{Rank-1 Tensor}
\label{apendice_polarizacion}
The solutions given in (\ref{eq:sol_hyper_vec}) must satisfy the first-order equations (\ref{eq:eps_vector}), in particular the $\xi_z$ component fulfills the equation
\begin{equation}
\label{eq:app_cero}
    k_{+}\partial_-\xi_z - k_-\partial_+\xi_z= 0 \,.
\end{equation}
The derivatives of $\xi_z$ are given in terms of $\xi_{\pm}$ and they can be found in the $\pm$ components of the first order equation (\ref{eq:eps_vector}). Writing them in terms of $\xi_1$ and $\xi_2$ leads to the equation

\begin{equation}
    \left(-2\beta\bar{k} - 2\beta i k_{+}(1-z)   -2\bar{k}z(1-z)\dfrac{d}{dz}\right)\xi_1 = \left(-2\beta k + 2\beta i k_{+}(1-z) + 2 k z(1-z)\dfrac{d}{dz}\right)\xi_2 \,.
    \label{eq:app_polarizacion}
\end{equation}
where $k=k_-+ik_+$ and $\bar{k}$ its complex conjugate. The equation above relates the constants $A$ and $B$. To solve it we can  use the following properties of the hypergeometric functions,
\begin{align}
	\label{app:prop_hyp_1}
	z(1-z)\dfrac{d}{dz}F(a,b,c,z) &= (1-z)b\left(F(a,b+1,c,z) - F(a,b,c,z)\right) \,, \\
	\label{app:prop_hyp_2}
	z(1-z)\dfrac{d}{dz}F(a+1,b+1,c,z) &= (c-a-1)F(a,b+1,c,z) \notag \\
										& \qquad+ (a+1-c+(b+1)z)F(a+1,b+1,c,z) \,, \\
	\label{app:prop_hyp_3}    
	a(1-z)F(a+1,b+1,c,z) &= (c-b-1)F(a,b,c,z)-(c-a-b-1)F(a,b+1,c,z) \,.
\end{align}
The LHS of the equation (\ref{eq:app_polarizacion}) can be written with the help of (\ref{app:prop_hyp_1}) as
\begin{equation}
   LHS= -2A b z^{\alpha}(1-z)^{\beta+1}\left[ik_+F(a,b,c,z)+\bar{k}F(a,b+1,c,z)\right] \,.
\end{equation}
For the RHS of (\ref{eq:app_polarizacion}) we use (\ref{app:prop_hyp_2}) first and then (\ref{app:prop_hyp_3}) to obtain
\begin{equation}
    \begin{aligned}
    RHS&=2Bz^{\alpha}(1-z)^{\beta+1}\left[ik_+a(1-z)F(a+1,b+1,c,z) + k(c-a-1)F(a,b+1,c,z)\right]\\
       &=2B(c-b-1)z^{\alpha}(1-z)^{\beta+1} \left[ik_+F(a,b,c,z)+\bar{k}F(a,b+1,c,z)\right] \,,
    \end{aligned}
\end{equation}
which means $A$ and $B$ satisfy

\begin{equation}
    \label{eq:ap_polariz}
    \dfrac{B}{A}= -\dfrac{c-b-1}{b} = \frac{2\beta-k}{2\beta+\bar{k}} \,.
\end{equation}

\subsection{Rank-2 Symmetric Tensor}
\label{apendice_polarizacion_2}
The boundary components of the first order equation (\ref{eq:epsilon_h}) in this case are
\begin{equation}
\label{eq:h_apendice}
\begin{split}
    -mh_{--}&=-h_{+-} + 2z(1-z)\left(\partial_zh_{+-}- \partial_+h_{z-}\right) \,, \\
    -mh_{++}&=h_{+-} + 2(1-z)\left(\partial_-h_{z+}-\partial_{z}h_{z+}\right) \,, \\
    -mh_{+-}&=h_{--} + 2z(1-z)\left(\partial_z h_{++} - \partial_{+}h_{z+}\right) \\
            &=-h_{++} + 2(1-z) \left(\partial_{-}h_{z-} - \partial_{z}h_{--}\right) \,,
\end{split}
\end{equation}
where the transversality condition for $h_{\mu\nu}$ was used to write down the last equation. The $h_{z+}$ and $h_{z-}$ component must fulfill equations similar to (\ref{eq:app_cero})

\begin{equation}
\begin{aligned}
    0=2z(1-z)\left(k_-\partial_+h_{z-}- k_+\partial_-h_{z-}\right) \,, \\
    0=2z(1-z)\left(k_+\partial_-h_{z+}- k_-\partial_+h_{z+}\right) \,.
\end{aligned}
\end{equation}
Using the relations (\ref{eq:h_apendice}) we can write this in terms of the following matrix equation
\begin{equation}
    0=\left[\begin{pmatrix}
    k_-m & -k_- +k_+mz & -zk_+\\
    k_- & k_+z+k_-m  &k_+mz
    \end{pmatrix} +
    2z(1-z)\begin{pmatrix}-k_+\partial_z & k_-\partial_z  & 0  \\ 0& -k_+ \partial_z & k_-\partial_z \end{pmatrix} \right] \begin{pmatrix} h_{--}\\ h_{+-}\\h_{++} \end{pmatrix} \,.
    \label{eq:app_pol_h}
\end{equation}
In order to write this in terms of the solutions of the hypergeometric equation $h_0,\, h_-\, h_+$ listed in (\ref{eq:sol_hyper_h}) we need to invert the equation (\ref{eq:cambio_base_h}), giving
\begin{equation}
    \begin{pmatrix} h_{--}\\ h_{+-}\\h_{++} \end{pmatrix} = \begin{pmatrix}1 &-1 & -1\\ 0 &-i &i \\1&1&1\\  \end{pmatrix} \begin{pmatrix} h_{0}\\ h_{-}\\h_{+} \end{pmatrix} \,.
\end{equation}
After using this, the easiest way to write down equations that only contain two of the functions $h_0,\, h_-\, h_+$ is to multiply the equation (\ref{eq:app_pol_h}) by the matrix
\begin{equation}
    c=\begin{pmatrix}
    1&-i\\1&i
    \end{pmatrix} \,,
\end{equation}
whose result is
\begin{equation}
\label{eq:eqhaches}
    0=\begin{pmatrix}a_{11} & a_{12} & 0 \\ a_{21} & 0 & a_{23} \end{pmatrix}\begin{pmatrix} h_{0}\\ h_{-}\\h_{+} \end{pmatrix} \,.
\end{equation}
Then, the second component of this equation involves $h_+$ and $h_0$ only, and it can be written as
\begin{equation}
    2\left(-2\beta\bar{k} - 2\beta ik_+(1-z) - 2\bar{k}z(1-z)\dfrac{d}{dz}\right)h_+ = \left(-2\beta k + 2\beta ik_+(1-z) + 2kz(1-z)\dfrac{d}{dz}\right)h_0 \,,
\end{equation}
which is the same equation as in the vector case (\ref{eq:app_polarizacion}). Therefore, by (\ref{eq:ap_polariz}) $A_0$ and $A_+$ satisfy
\begin{equation}
    \dfrac{A_0}{2A_+}= \dfrac{2\beta -k }{2\beta + \bar{k}}=\dfrac{im-1 - (k_-+ik_+)}{im-1+(k_--ik_+)} \,.
\end{equation}
The first component of the matrix equation \eqref{eq:eqhaches} can also be put in the form of (\ref{eq:app_polarizacion}) and then
\begin{equation}
    \dfrac{2A_-}{A_0}= \dfrac{2(\beta+1)-k}{2(\beta+1)+\bar{k}} =\dfrac{im+1-(k_-+ik_+)}{im+1 + (k_--ik_+)} \,.
\end{equation}

\subsection{Asymptotic Behaviour}
\label{apendice_asintotica}
We want to analyze the behavior of functions of the form
\begin{equation}
    R(z)=z^{\alpha}(1-z)^{\beta}F(a,b,c,z) \,,
\end{equation}
near $z\sim 1$, where the constants $\alpha,\, \beta,\, a,\, b$ and $c$ satisfy the relations (\ref{eq:alpha_beta}) and (\ref{eq:abc}). First, let us suppose that $a+b-c=2\beta-1$ is not an integer or zero\footnote{The same procedure can be done with help of the functions $\textbf{F}(a,b,c,z)=F(a,b,c,z)/\Gamma(c)$ which are well-defined solutions of the hypergeometric differential equation for all values of $a$, $b$ and $c$. A more general connection formula can be written in terms of $\textbf{F}(a,b,c,z)$. Such formula reduces to (\ref{eq:conecction_formula}) when $a+b-c$ is not an integer or zero \cite{Olver}. }, then we can use the connection formula

\begin{equation}
    \begin{aligned}
        F(a,b,c,z)=&\dfrac{\Gamma(c)\Gamma(c-a-b)}{\Gamma(c-a)\Gamma(c-b)}F(a,b,1+a+b-c,1-z)\\
        &+ \dfrac{\Gamma(c)\Gamma(a+b-c)}{\Gamma(a)\Gamma(b)}(1-z)^{c-a-b}F(c-a,c-b,1+c-a-b,1-z) \,.
    \end{aligned}
    \label{eq:conecction_formula}
\end{equation}
Performing a change of variables $z=\tanh^2(\varZ)$ and using the connection formula we obtain the following expression
\begin{equation}
    \begin{aligned}
    R(\varZ)= \tanh^2(\varZ)&\left[ \dfrac{\Gamma(c)\Gamma(c-a-b)}{\Gamma(c-a)\Gamma(c-b)}\dfrac{1}{\cosh^{2\beta}(\varZ)}F(a,b,1+a+b-c,1/\cosh^2(\varZ)) \right. \\
    & \left.  +\dfrac{\Gamma(c)\Gamma(a+b-c)}{\Gamma(a)\Gamma(b)}\dfrac{1}{\cosh^{2(1-\beta)}(\varZ)} F(c-a,c-b,1+c-a-b,1/\cosh^2(\varZ)) \right] \,,
    \end{aligned}
\end{equation}
for which it is easier to evaluate its behavior when $\varZ\sim \infty$
\begin{equation}
    \begin{aligned}
    R(\varZ)\sim \dfrac{\Gamma(c)\Gamma(c-a-b)}{\Gamma(c-a)\Gamma(c-b)}e^{-2\beta\varZ} \left(1+\mathcal{O}(e^{-2\varZ})\right) +\dfrac{\Gamma(c)\Gamma(a+b-c)}{\Gamma(a)\Gamma(b)} e^{2(\beta-1)\varZ} \left(1+\mathcal{O}(e^{-2\varZ})\right) \,.
    \end{aligned}
    \label{app:asintotica}
\end{equation}
It is clear that if $\beta$ has a real part, one of the exponential is divergent and it may cause problems in the norm of the tensors.

\newpage
\bibliographystyle{toine}
\bibliography{3Dgravitybib}{}

\end{document}